\documentclass[preprint]{revtex4}

\newcommand{\be}{\begin{equation}}
\newcommand{\ee}{\end{equation}}
\newcommand{\bea}{\begin{eqnarray}}
\newcommand{\eea}{\end{eqnarray}}

\begin{document}

\preprint{SU-GP-03/2-2}
 
\setlength{\unitlength}{1mm}

\title{Baryogenesis and the New Cosmology}

\author{Mark Trodden}
\affiliation{Department of Physics, Syracuse University,
Syracuse, NY 13244-1130, USA}

\begin{abstract}In this talk I begin with a brief review of the status of approaches to understanding the origin of the baryon asymmetry of the universe (BAU). I then describe a recent model unifying three seemingly-distict problems facing particle cosmology: the origin of inflation, the generation of the BAU and the nature of dark energy.
\vspace{3cm}
\begin{center}
{\it Plenary talk presented at PASCOS-03, Mumbai, India; COSMO-02, Chicago, and at  the Aspen Winter 2003 Conference on Particle Physics: At the Frontiers of Particle Physics, Aspen Center for Physics. To appear in the proceedings of PASCOS-03.}
\end{center}
\end{abstract}

\maketitle

\section{Introduction}
The audience for this talk was extremely diverse, ranging from beginning graduate students, through experts in subfields of physics somewhat distinct from the subject matter of my talk, to baryogenesis experts. My strategy therefore was to present an overview of the main issues facing baryogenesis, and only to focus on the more technical aspects of a particular model towards the end of the talk.  Much of the review portion of these proceedings draws heavily on my article with Riotto~\cite{Riotto:1999yt} and  the latter parts summarize my recent paper with De Felice and Nasri~\cite{DeFelice:2002ir}.

The problem of the baryon asymmetry of the universe is a classic problem of particle cosmology. Particle physics has taught us that matter and antimatter behave essentially identically, and indeed the interactions between matter and antimatter are the focus of successful terrestrial experiments. On the other hand, cosmology teaches us that the early universe was an extremely hot, and hence energetic, environment in which one would expect equal numbers of baryons and antibaryons to be copiously produced. This early state of the universe stands in stark contrast to what we observe in the universe today. Direct observation shows that the universe around us contains no appreciable primordial antimatter. In addition, the stunning success of big bang nucleosynthesis rests on the requirement that, defining $n_{b({\bar b})}$ to be the number density of (anti)-baryons and $s$ to be the entropy density,
\begin{equation}
2.6\times 10^{-10} < \eta\equiv \frac{n_b -n_{\bar b}}{s} < 6.2\times 10^{-10} \ ,
\end{equation}
(see, for example,~\cite{Fields:cn}).
Very recently this number has been independently determined to be $\eta =  6.1\times 10^{-10}\ ^{+0.3\times 10^{-10}}_{-0.2\times 10^{-10}}$ from precise measurements of the relative heights of the first two microwave background (CMB) acoustic peaks by the WMAP satellite~\cite{Bennett:2003bz}.
Thus the natural question arises; as the universe cooled from early times to today, what processes, both particle physics and cosmological, were responsible for the generation of this very specific baryon asymmetry?

The outline of this talk is as follows. In the next section I survey several mechanisms by which the BAU might be generated. In particular I focus on electroweak baryogenesis and discuss why it is, to my
mind, a particularly attractive mechanism. In section~\ref{quint} I will then describe a model which attempts to unify three seemingly-disparate problems of modern cosmology; the origin of inflation, the generation of the BAU and the nature of dark energy.

\section{Our Favorite Models}
\label{favorites}
If we're going to use a particle physics model to generate the baryon asymmetry
of the universe (BAU), what properties must the theory possess? This question
was first addressed by Sakharov~\cite{Sakharov:dj} in 1967, resulting in the
following criteria

\begin{itemize}
\item Violation of the baryon number ($B$) symmetry.
\item Violation of the discrete symmetries $C$ (charge conjugation)
      and $CP$ (the composition of parity and $C$)
\item A departure from thermal equilibrium.
\end{itemize}
Of course, the first of these is obvious - no $B$ violation, no baryon production.
To understand the second condition, note that, roughly speaking, 
if $C$ and $CP$ are conserved, the
rate for any process which generates baryons is equal to that for the conjugate 
process, which produces antibaryons, so no net excess is generated on average.
Finally, in thermal equilibrium the number density of a particle species is
determined purely by its energy, and since the masses of particle and antiparticle
are equal by the CPT theorem, the number density of baryons equals that of
antibaryons. We will want to recall this connection between the equilibrium condition and the CPT theorem in the last part of this talk.

Grand Unified baryogenesis (for a review see ~\cite{Langacker:1980js})
was the first implementation of Sakharov's
baryon number generation ideas. There are at least three reasons why this is
an attractive mechanism. First, baryon number violation is an unavoidable
consequence of Grand Unification. If one wishes to have quarks and leptons
transforming in the same representation of a single group, then there
necessarily exist processes in which quarks convert to leptons and vice-versa.
Hence, baryon number violation is manifest in these theories. Second, it is
a simple and natural task to incorporate sufficient amounts of CP-violation
into GUTs. Among other possibilities, there exist many possible mixing phases
that make this possible. Finally, at such early epochs of cosmic evolution,
the relevant processes are naturally out of equilibrium since the relevant
timescales are slow compared to the expansion rate of the universe. The
simplicity with which the Sakharov criteria are achieved in GUTs makes this
mechanism very attractive.

While GUT baryogenesis is attractive, it is not likely that the physics 
involved will be directly testable in the foreseeable future. While we may gain
indirect evidence of grand unification with a particular gauge group, direct
confirmation in colliders seems unrealistic. A second problem with GUT
scenarios is the issue of erasure of the asymmetry. As we will see, the
electroweak theory ensures that there are copious baryon number violating
processes between the GUT and electroweak scales. These events
violate $B+L$ but conserve $B-L$. Thus, unless a GUT mechanism generates an
excess $B-L$, any baryonic asymetry produced will be equilibrated to zero
by anomalous electroweak interactions. While this does not invalidate GUT 
scenarios, it is a constraint. For example, $SU(5)$ will not be suitable
for baryogenesis for this reason, while $SO(10)$ may be. This makes  the idea
of baryogenesis through leptogenesis particularly attractive. 

Affleck-Dine baryogenesis~\cite{Affleck:1984fy} is a particularly attractive scenario, and much
progress has been made in understanding how this mechanism works. As was the
case for electroweak baryogenesis, this scenario has found its most promising 
implementations~\cite{Dine:1995uk,Dine:1995kz} in supersymmetric models, in which the necessary flat 
directions are abundant. Particularly attractive is the fact that these
moduli, carrying the correct quantum numbers, are present even in the MSSM.

The challenges faced by Affleck-Dine models are combinations of those faced
by the GUT and electroweak ideas. In particular, it is necessary that $B-L$
be violated along the relevant directions (except perhaps in the Q-ball
implementations~\cite{Kusenko:1997zq}) and that there exist new physics at scales above the
electroweak. If supersymmetry is not found, then it is hard to imagine
how the appropriate flat directions can exist in the low energy models.

In recent years, perhaps the most widely studied scenario for generating 
the baryon number of the universe has been electroweak baryogenesis. The
popularity of this idea is tightly bound to its testability. The physics
involved is all testable in principle at realistic colliders. Furthermore,
the small extensions of the model involved to make baryogenesis successful
can be found in supersymmetry, which is an independently attractive
idea, although electroweak baryogenesis does not depend on supersymmetry. I will now
describe this briefly.

\subsection{Baryon Number Violation in the EW Theory}
In the standard electroweak theory baryon number is an exact global symmetry.
However, baryon number is violated
at the quantum level through nonperturbative processes. These effects are
closely related to the nontrivial vacuum structure of the electroweak theory.

At zero temperature, baryon number violating events are exponentially suppressed.
This is because there exists a potential barrier between vacua and anomalous
processes are thus tunneling events. The relevant barrier height is set by the
point of least energy on the barrier. This point is known as the 
{\it sphaleron}~\cite{Klinkhamer:1984di}, and has energy $E_{sph}\sim 10$ TeV.
However, at temperatures above or comparable to the critical temperature
of the electroweak phase transition, vacuum transitions over the sphaleron 
may occur frequently due to thermal activation~\cite{Kuzmin:1985mm}.

\subsection{C and CP Violation in the EW Theory}
Fermions in the electroweak theory are chirally coupled to the gauge fields. 
In terms of the discrete symmetries of the theory,
these chiral couplings result in the electroweak theory being maximally
C-violating.
However, the issue of CP-violation is more complex.

CP is known not to be an exact symmetry
of the weak interactions, and is observed experimentally in the neutral 
Kaon system through $K_0$, ${\bar K}_0$ mixing. Although at
present there is no completely satisfactory theoretical explanation 
of this, CP violation is a natural feature of the
standard electroweak model. The Kobayashi-Maskawa (KM) quark mass mixing
matrix contains a single
independent phase, a nonzero value for which signals CP violation.
While this is encouraging for baryogenesis, it turns out that this particular source of
CP violation is not strong enough. The relevant effects are parametrized by
a dimensionless constant which is no larger than $10^{-20}$. This appears
to be much too small to account for the observed BAU and so it is usual to turn
to extensions of the minimal theory. In particular the minimal supersymmetric standard
model (MSSM).

\subsection{The Electroweak Phase Transition}
The question of the order of the electroweak phase transition is central to
electroweak baryogenesis. Since the equilibrium description of particle 
phenomena is extremely accurate at electroweak temperatures, baryogenesis 
cannot occur at such low scales without the aid of phase transitions.

For a continuous transition, the associated departure from
equilibrium is insufficient to lead to relevant baryon number production. 
The order parameter for the electroweak phase transition is
$\varphi$, the modulus of the Higgs field.
For a first order transition the extremum at $\varphi=0$ becomes separated
from a second local minimum by an energy barrier.
At the critical temperature $T=T_c$ both phases are equally 
favored energetically and at later times the minimum at $\varphi \neq 0$ becomes
the global minimum of the theory. Around $T_c$ quantum tunneling
occurs and nucleation of bubbles of the true vacuum 
in the sea of false begins. At a particular temperature below $T_c$, bubbles
just large enough to grow nucleate. These are termed {\it critical} bubbles,
and they expand, eventually filling all of space and completing the transition.
As the bubble walls pass each point in space, the order
parameter changes rapidly, as do the other fields and this leads to a
significant departure from thermal equilibrium. Thus, if the phase 
transition is strongly enough first order it is possible to satisfy
the third Sakharov criterion in this way.

There is a further criterion to be satisfied. As the wall passes a
point in space, the Higgs fields evolve rapidly and the Higgs VEV changes from
$\langle\phi\rangle=0$ in the unbroken phase to
\be
\langle\phi\rangle=v(T_c)
\label{vatTc}
\ee
in the broken phase. Here, $v(T)$ is the value
of the order parameter at the symmetry breaking global minimum of the finite 
temperature effective potential. 
Now, CP violation and the departure from equilibrium occur while the Higgs field 
is changing. Afterwards, the point is
in the true vacuum, baryogenesis has ended, and baryon number violation
is exponentially supressed. Since baryogenesis is now over, 
it is
imperative that baryon number violation be negligible at this temperature in
the broken phase, otherwise any baryonic excess generated will be
equilibrated to zero. Such an effect is known as {\it washout} of the 
asymmetry and the criterion for this not to happen may be written as
\be
\frac{v(T_c)}{T_c} \geq 1 \ .
\label{washout}
\ee
Although there are a number of nontrivial steps
that lead to this simple criterion, (\ref{washout}) is traditionally used 
to ensure that the baryon asymmetry survives after
the wall has passed.
It is necessary that this criterion be satisfied for any electroweak 
baryogenesis scenario to be successful.

In the minimal standard model, in which experiments at LEP now constrain the Higgs mass 
to be $m_H > 114.4$ GeV, it is clear from numerical simulations
that (\ref{washout}) is not satisfied. This is therefore a second
reason to turn to extensions of the minimal model.

In the MSSM there are two Higgs fields, $\Phi_1$ and $\Phi_2$. At one loop, a CP-violating
interaction between these fields is induced through supersymmetry
breaking. Alternatively, there also exists extra CP-violation through
CKM-like effects in the chargino mixing matrix. Thus, there seems to be
sufficient CP violation for baryogenesis to succeed.

Now, the two Higgs fields combine to give one lightest scalar Higgs $h$. 
In addition, there are also light {\it stops} ${\tilde t}$ (the
superpartners of the top quark) in the theory. These light scalar
particles can lead to a strongly first order phase transition if the 
scalars have masses in the correct region of parameter space. A detailed
two loop calculation~\cite{Carena:2002ss} and lattice results indicate that the allowed region is given by
\bea
m_h & \leq 120 {\rm GeV} \\
m_{\tilde t} & \leq m_t \ ,
\label{MSSMconstraints}
\eea  
for $\tan\beta \equiv \langle \Phi_2 \rangle/\langle \Phi_1 \rangle > 5$.
In the next few years, experiments at the Tevatron and the LHC should probe this range of Higgs 
masses and we should know if the MSSM is a good candidate for electroweak
baryogenesis.

Thus, the testability of electroweak scenarios also leads to tight 
constraints, and at present there exists only a small window of parameter space in
extensions of the electroweak theory in which baryogenesis is viable. 

If the Higgs is found, the second
test will come from the search for the lightest stop at the Tevatron
collider. If both particles are found, the last crucial test will
come from CP-violating effects which may be observable in $B$ physics.
Moreover, the preferred parameter space leads
to values of the branching ratio ${\rm BR}(b\rightarrow s\gamma)$
different from the Standard Model case. Although the exact value
of this branching ratio depends strongly on the value of the $\mu$
and $A_t$ parameters, the typical difference with respect to the
Standard
Model prediction is of the order of the present experimental
sensitivity
and hence in principle testable in the near future. Indeed, for the
typical spectrum considered here, due to the light charged Higgs,
the
branching ratio ${\rm BR}(b \rightarrow s \gamma)$ is somewhat
higher than in the SM case, unless
negative values of $A_t\mu$  are
present. 

Having given an overview of baryogenesis and a description of a particularly attractive mechanism, I will turn, in the rest of my talk, to a mechanism I've been working on recently - {\it quintessential baryogenesis}~\cite{DeFelice:2002ir}.

\section{Quintessential Baryogenesis}
\label{quint}
Rolling scalar fields are a mainstay of modern cosmology. This is perhaps best-illustrated by the 
inflationary paradigm~\cite{Guth:1980zm,Linde:1981mu,Albrecht:1982wi}, in which most implementations involve a scalar field rolling towards the
minimum of its potential in such a way that the potential energy of the field is the dominant
component of the energy density of the universe. There are, however, many other cosmological
instances in which scalar fields are invoked. In fact, it is often felt that cosmologists are prepared to invent a new scalar field every time a new piece of cosmological data comes along.

During the last few years a new consistent picture of the energy budget of the universe has emerged.
Large scale structure studies show that matter (both luminous and dark) contributes a fraction of
about 0.3 of the critical density, while the position of the first acoustic peak of the cosmic microwave
background power spectrum indicates that the total energy density is consistent with criticality. The
discrepancy between these two measurements may be reconciled by invoking a negative pressure 
component
which is termed {\it dark energy} and leads to the acceleration of the universe.
Current limits~\cite{Melchiorri:2002ux}, obtained by combining results 
from cosmic microwave background experiments with large scale structure data, 
the Hubble parameter measurement from the Hubble Space Telescope and luminosity measurements of Type Ia supernovae, give $-1.62< w <-0.74$ at the $95 \%$ 
confidence level.

It is of course possible that this mystery component is a cosmological constant $\Lambda$, for
which $w_{\Lambda}=-1$. Alternatively, it has been suggested~\cite{Wetterich:fm}-\cite{Caldwell:1997ii}
that if the cosmological constant itself is zero,
the dark energy component could be due to the dynamics of a rolling scalar field, in a form of
late-universe inflation that has become known as {\it quintessence}. 

It is natural to wonder whether the inflaton and the quintessence field might be one and the 
same~\cite{Spokoiny:1993kt}, and, in fact, specific models for this have been 
proposed~\cite{Spokoiny:1993kt}-\cite{Dimopoulos:2001ix}.
In this part of my talk I'll investigate how we may further limit the proliferation of rolling scalar fields
required in modern cosmology by studying how a scalar field responsible both for inflation and 
for late-time acceleration of the universe might also be responsible for the generation of the baryon asymmetry of the universe. The relationship between early-time acceleration -- inflation -- and baryogenesis has been 
explored in some detail (for example see~\cite{Affleck:1984fy},~\cite{Yokoyama:1986gx}-\cite{Nanopoulos:2001yu}).
Here I consider the opposite regime, that the quintessence field may be associated with the
generation of the baryon asymmetry~\cite{DeFelice:2002ir} (for interesting related studies see~\cite{Li:2001st,Feng:2002nb,Brandenberger:2003kc}). 

\subsection{Quintessential Inflation} 
I'll use the flat Friedmann, Robertson-Walker (FRW) ansatz for the metric
\begin{equation}
\label{metric}
ds^2=-dt^2+a(t)^2\left[dr^2+r^2d\Omega_2^2\right] \ .
\end{equation}
Here the energy density $\rho$ and pressure $p$ for a real homogeneous scalar field $\phi$
are given by
\begin{equation}
\label{energydensity}
\rho_{\phi}=\frac{1}{2}{\dot \phi}^2+V(\phi) \ ,
\end{equation}
\begin{equation}
\label{pressure}
p_{\phi}=\frac{1}{2}{\dot \phi}^2-V(\phi) \ ,
\end{equation}
respectively, with $V(\phi)$ the potential, and where we have defined the Planck mass by $G\equiv M_{\rm p}^{-2}$. The scalar field
itself obeys
\begin{equation}
\label{phieqn}
{\ddot \phi}+3\left(\frac{{\dot a}}{a}\right){\dot \phi}+\frac{dV(\phi)}{d\phi}=0 \ ,
\end{equation}
with a prime denoting a derivative with respect to $\phi$.

Now, to explain the current data indicating an accelerating universe, it is necessary to have the
dominant type of matter at late times be such that ${\ddot a}>0$. If this matter is to be $\phi$, 
then this implies $\rho_{\phi} +3p_{\phi} <0$. Writing $p_{\phi}\equiv w_{\phi}\rho_{\phi}$ this translates into an equation of state parameter that obeys $w_{\phi}<-\frac{1}{3}$

In the Peebles and Vilenkin model~\cite{Peebles:1998qn} one uses the potential
\begin{equation}
V(\phi)=\left\{\begin{array}{lll}
\lambda (\phi^4 +M^4) & \ \ \ \ \ , \ \ \ \ \ \ & \phi\in (-\infty,0] \\
\frac{\lambda M^8}{\phi^4 +M^4} & \ \ \ \ \ , \ \ \ \ \ \ & \phi\in (0,\infty) \ ,
\end{array}\right.
\end{equation}
We require inflation to occur at early times,
and indeed, for this potential, the slow-roll conditions are satisfied 
for sufficiently large and negative $\phi$.

Inflation ends when the slow-roll conditions are violated and the potential and
kinetic energies of the inflaton are comparable with each other. 
In traditional inflationary models the inflaton then rapidly transfers its energy to other fields either through perturbative effects (reheating) or parametric resonance (preheating). Here, however,
there is no such effect, and it is the kinetic energy of the field $\phi$ that is the dominant component
of the energy density of the universe immediately after the end of inflation. Following 
Joyce~\cite{Joyce:1996cp} we term this behavior {\it kination}.

A successful cosmology requires the universe be radiation-dominated at the time of nucleosynthesis,
since otherwise the precision predictions of that theory are no longer in agreement with
observations. The lack of conventional reheating, the conversion of the potential energy of the
inflaton to particle production, in quintessential inflation means that the requisite radiation must
be produced another way. In fact, the radiation era in these models is due to the subtle behavior
of quantum fields in changing geometries.

At the end of inflation, the FRW line element undergoes an abrupt change from that associated with
cosmic expansion (exponential or power-law) to that associated with kination. Massless quantum
fields in their vacua in the inflation era are no longer in vacuum in the kination era, corresponding
to gravitational particle production. This effect is analogous to Hawking radiation, and has been explored
in detail~\cite{Ford:1986sy}-\cite{Giovannini:1998bp} in the cosmological context of interest here.

For a significant 
time subsequent to this, cosmic evolution is much the same as in the standard cosmology, with
a matter dominated epoch eventually succeeding the radiation era. Although the scalar field is not
important during these times, the density fluctuations seeded by quantum fluctuations in $\phi$ 
during inflation lead to structure formation and temperature fluctuations in the cosmic microwave 
background radiation. 

Finally, consider this extreme future of the universe, in which the scalar field becomes responsible for quintessence. It is clear that when quintessence begins, the energy density of the field $\phi$ once again
becomes dominated by its potential energy density. 
It is a challenge similar to that for conventional quintessence to ensure that this epoch occurs at
the present time and yields the correct ratio of matter to dark energy.

\subsection{Generation of the Baryon Asymmetry}
\label{qbg}
In order for the quintessence field $\phi$ to play a role in baryogenesis, we must consider how $\phi$
couples to other fields. In principle, the inflaton and quintessence field may lie in any sector of the 
theory, the phenomenologically safest of which would be one in which there are only gravitational 
strength
couplings to other particles. Here I adopt a conservative approach
and assume that $\phi$ couples to standard model fields with couplings specified by a dimensionless
constant and an energy scale which we shall leave as a free parameter for the moment and later 
constrain by observations and the condition that our model produce a sufficient baryon asymmetry.

Consider terms in the effective Lagrangian density of the form
\begin{equation}
\label{jmucoupling}
{\cal L}_{\rm eff}=\frac{\lambda'}{M}\partial_{\mu}\phi J^{\mu} \ ,
\end{equation}
where $\lambda'$ is a coupling constant, $M<M_{\rm p}$ is the scale of the cutoff in the effective theory
and $J^{\mu}$ is the current corresponding to some continuous global symmetry such as baryon
number or baryon number minus lepton number. Further, let us
assume that $\phi$ is homogeneous. We then obtain
\begin{equation}
\label{coupling}
{\cal L}_{\rm eff}=\frac{\lambda'}{M}{\dot \phi}\ \Delta n \equiv \mu(t)\Delta n \ ,
\end{equation}
where $n=J^0$ is the number density corresponding to the global symmetry and we have
defined an effective time-dependent ``chemical potential'' $\mu(t)\equiv \lambda'{\dot \phi}/M$. 

Recall that we need to satisfy the Sakharov criteria in order to generate a baryon asymmetry 
(for reviews see~\cite{Cohen:1993nk}-\cite{Trodden:1998ym},~\cite{Riotto:1999yt}). 
The first of these requires baryon number $B$ to be violated. At this stage, to maintain 
generality, we shall leave the mechanism of baryon number violation unspecified. Further, the standard
model is maximally C-violating due to its chiral structure, and the coupling~(\ref{jmucoupling}) is
$CP$-odd. In this sense, no {\it explicit} $CP$-violation is required in this model.
The third Sakharov criterion requires a departure from thermal equilibrium if $CPT$ is a manifest symmetry. However, the crucial point about baryogenesis in the presence of the rolling scalar
field $\phi$ is that $CPT$ 
is broken spontaneously by the explicit value taken by $\langle{\dot \phi}\rangle \neq 0$.
Thus, the particular model of baryogenesis that is important here is 
{\it spontaneous baryogenesis}~\cite{Cohen:1988kt},
which is effective even in thermal equilibrium.
Following~\cite{DeFelice:2002ir}, I will refer to this model, in which the rolling scalar responsible for both inflation and dark energy
also provides a source for spontaneous baryogenesis as {\it quintessential baryogenesis}.

To understand how spontaneous (and hence quintessential) baryogenesis works, note that in thermal equilibrium we have
\begin{equation}
\Delta n(T;\xi)=\int \frac{d^3{\bf p}}{(2\pi)^3}[f(E,\mu)-f(E,-\mu)] \ ,
\end{equation}
where $\xi\equiv \mu/T$ is a parameter and $f(E,\mu)$ is the phase-space distribution of the particles
of the current $J^{\mu}$, which may be Fermi-Dirac or Bose-Einstein. Thus, for $\xi <1$
\begin{equation}
\Delta n(T;\mu)\simeq \frac{gT^3}{6} \xi+{\cal O}(\xi^2) \ ,
\label{deltan}
\end{equation}
where $g$ is the number of degrees of freedom of the field corresponding to $n$. Therefore,
\begin{equation}
\Delta n(T;\mu)\simeq \frac{\lambda' g}{6M} T^2{\dot \phi} \ .
\end{equation}

Now recall that the entropy density is given by
\begin{equation}
s=\frac{2\pi}{45} g_* T^3 \ ,
\end{equation}
where $g_*$ is the effective number of relativistic degrees of freedom in thermal equilibrium at
temperature $T$. Whatever the mechanism of baryon number violation, there will exist a temperature
$T_F$ below which baryon number violating processes due to this mechanism become sufficiently
rare that they freeze out. For $T<T_F$ these processes can no longer appreciably change the baryon 
number of the universe. Computing the freeze-out value of the baryon to entropy ratio we then obtain
\begin{equation}
\label{baueqn}
\eta_F\equiv \eta(T_F) \equiv \frac{\Delta n}{s}(T_F) \simeq 0.38 \lambda' \left(\frac{g}{g_*}\right)\frac{{\dot \phi}(T_F)}{MT_F} \ .
\end{equation}

How the baryon excess evolves after this point depends on the value of $T_F$ and on the
relevant current in equation~(\ref{jmucoupling}). If 
$T_F \leq T_c^{\rm EW}\sim 100$ GeV, the critical temperature of the electroweak phase transition,
then all baryon number violation ceases at $T_F$ and $\eta(T<T_F)=\eta_F$. However, if
$T_F>T_c^{\rm EW}$, then we must take into account the effects of anomalous electroweak
processes at finite temperature. These can be involved in directly generating the baryon asymmetry
(electroweak baryogenesis), in reprocessing an asymmetry in other quantum numbers into one in
baryon number (for example in leptogenesis) or in diluting the asymmetry created by any 
baryogenesis mechanism which is effective above the electroweak scale and does not produce a
$B-L$ asymmetry. It is important to realize that, in the context of quintessential inflation, the
quantitative effects of these electroweak processes may differ substantially from those in the standard 
cosmology~\cite{Joyce:1996cp,Joyce:1997fc} since in our case, 
the electroweak phase transition may occur during kination rather than radiation domination.

I have now provided quite a general description of quintessential baryogenesis, appropriate to the level of this talk. While this allows us to demonstrate the generic features of our model, we cannot calculate the magnitude of
the actual baryon asymmetry generated without first specifying a mechanism of baryon number
violation (and hence a value for $T_F$) and a value for the dimensionless combination 
$\lambda' M_{\rm p}/M$. Here I'll provide just one concrete example.

\subsection{Baryon Number Violation Through Non-renormalizable Operators}
If there exists baryon number violating physics above the standard model, then this physics
will manifest itself in non-renormalizable operators  in the standard model. For the purposes of this
section we will actually be interested in operators that violate the anomaly-free combination
$B-L$. Consider the effective 4-fermion operator
\begin{equation}
\label{4fermion}
{\cal L}_{B-L}=\frac{{\tilde g}}{M_X^2} \psi_1\psi_2{\bar \psi}_3{\bar \psi}_4 \ ,
\end{equation}
where $\psi_i$ denote standard model fermions. Here ${\tilde g}$ is a dimensionless coupling, obtained after integrating out the $B-L$ violating effects of a particle of mass $M_X$. 
The rate of baryon number violating processes due to this operator is, as usual, defined by
$\Gamma_{B-L}(T)=\langle \sigma(T) n(T) v \rangle$, where $\sigma (T)$ is the cross-section for
$\psi_1 + \psi_2 \rightarrow \psi_3 + \psi_4$, $n(T)$ is the number density of $\psi$ particles, $v$
is the relative velocity and $\langle \cdots \rangle$ denotes a thermal average. 
For temperatures
$T<M_X$ we have $n(T)\sim T^3$, $\sigma (T) \sim {\tilde g}^2 T^2/M_X^4$, and $v\sim 1$ which 
yields
\begin{equation}
\Gamma_{B-L}(T)\simeq \frac{{\tilde g}^2}{M_X^4} T^5 \ .
\end{equation}
The high power of the temperature dependence in this rate results from the fact 
that~(\ref{4fermion}) is an irrelevant operator in the electroweak theory and is
crucial for the success of our mechanism.
These interactions are in thermal equilibrium in the early universe, but because their rate drops
off so quickly with the cosmic expansion they will drop out of equilibrium at the temperature $T_F$ 
defined through
\begin{equation}
\Gamma_{B-L}(T_F)=H(T_F) \ .
\end{equation}

As a definite example, let us take $\lambda' M_{\rm p}/M\sim 8$, which is as large as is allowed by
current constraints. In this case we need 
$M_X^{\rm PV}\sim 10^{11}$ GeV, the intermediate scale that appears in some supersymmetric 
models.

\subsection{Brief Comments on Constraints and Tests}
\label{sec:constraints}
The presence of an extremely light scalar field in the universe has the potential to lead to a number
of observable consequences in the laboratory and in cosmology. In the case of quintessence these
effects have been analyzed in some detail by Carroll~\cite{Carroll:1998zi}. Particularly strong constraints
arise due to couplings of the form
\begin{equation}
\label{phiFFdual}
{\cal L}_{\rm eff}^{(1)}\equiv \beta_{F{\tilde F}}\frac{\phi}{M}F_{\mu\nu}{\tilde F}^{\mu\nu} \ ,
\end{equation}
where $F_{\mu\nu}$ and 
${\tilde F}^{\mu\nu}\equiv \frac{1}{2}\epsilon^{\mu\nu\rho\sigma}F_{\rho\sigma}$ are the
electromagnetic field strength tensor and its dual respectively. If, as in quintessence models, the field
$\phi$ is homogeneous and time varying, then it affects the dispersion relation for electromagnetic
waves and leads to a rotation in the direction of polarized light from radio 
sources~\cite{Carroll:vb,Carroll:1991zs}. To avoid such bounds, quintessence models usually struggle to have such couplings be as small as
possible. In our model however, we are making important use of the coupling~(\ref{jmucoupling}). If
the relevant current is that for baryon number $J^{\mu}_B$, then it can be shown that Carroll's bound
yields $\lambda' M_{\rm p}/M < 8$ .
If the relevant current is $J^{\mu}_{B-L}$ then the above argument does not apply formally.
However, we may still generally expect an analogous coupling to the term 
$F^{\mu\nu} {\tilde F}_{\mu\nu}$ of a similar order.

We have already demonstrated that, for an appropriate scale $M_X$, 
successful quintessential baryogenesis takes place for 
$\lambda'  < 8$, and so it is possible to generate the observed BAU and to evade existing
constraints. If quintessential baryogenesis is correct, then it may be that the relevant coupling
$\lambda'$ lies just below the existing observational bounds and that future studies of the
rotation of polarized light from distant galaxies will reveal the presence of such a term.

\section{Comments and Conclusions}
\label{conclusions}
In the first part of this talk I presented a current view of the status of baryogenesis models, focusing on the best-known ones.

In this second part of my talk I extended the approach of Spokoiny~\cite{Spokoiny:1993kt}
and of Peebles and Vilenkin~\cite{Peebles:1998qn} in exploring the extent to which the dynamics
of a single scalar field can be responsible for not only inflation and dark
energy domination, but also the baryon asymmetry of the universe. This model is called quintessential baryogenesis~\cite{DeFelice:2002ir}, and is interesting, of course,
since the best fit cosmology to all current observational data is one in which the universe undergoes two
separate epochs of acceleration. 

We have left a number of questions unanswered and will return to them in future work. Perhaps the
most pressing issue is one that plagues rolling scalar models of dark energy in general, namely the
question of technical naturalness of the potentials involved, and their stability to quantum
corrections. However, this is a general issue for quintessence models, and is
not specific to our baryogenesis mechanism. 

One thing is certain: the existence of the BAU is clear evidence from cosmology of particle physics beyond the standard model. This makes the conundrum of the BAU a fascinating problem for cosmologists and high-energy physicists alike.

\section*{Acknowledgements}
I would like to thank Antonio De Felice and Salah Nasri for a very enjoyable collaboration on quintessential baryogenesis. I would like to thank the organizers of the PASCOS-03 conference at the TIFR in Mumbai for their hospitality and for a wonderful conference. This work is supported by the National Science Foundation (NSF) under grant PHY-0094122.

\end{document}